\documentclass[envcountsame, runningheads]{llncs}

\usepackage[utf8]{inputenc}
\usepackage[T1]{fontenc}
\usepackage{lmodern}
\usepackage[hidelinks]{hyperref}
\usepackage{diagbox}

\usepackage{amssymb}
\usepackage{amsmath}
\usepackage{stmaryrd}
\usepackage{tikz}
\usetikzlibrary{backgrounds, positioning, arrows, calc}
\usepackage{mathtools}
\usepackage[linesnumbered,vlined,noend, ruled]{algorithm2e}
\usepackage[inline]{enumitem}  
\usepackage{cleveref}
\usepackage{lineno}
\usepackage{booktabs}
\usepackage{multirow}
\renewcommand{\cref}{\Cref}

\usepackage{todonotes}

\newcommand{\DD}{\ensuremath{\mathbb{D}}\xspace}

\newcommand{\QQ}{\ensuremath{\mathbb{Q}}\xspace}
\newcommand{\RR}{\ensuremath{\mathbb{R}}\xspace}

\newcommand{\A}{\mathcal{A}}
\newcommand{\B}{\mathcal{B}}
\newcommand{\C}{\mathcal{C}}

\newcommand{\trans}[3]{#1\xrightarrow[]{#2}#3}

\tikzstyle{state}=[thick,minimum size=18pt, circle,draw]
\tikzstyle{transition}=[->,thick,>=stealth,shorten >=1pt,shorten <=1pt]
\tikzstyle{final}=[after node path={ node[state, scale=.8] at (\tikzlastnode) {} }]
\tikzstyle{initial}=[after node path={
	[to path={[transition] (\tikztostart) -- (\tikztotarget)}]
	(\tikzlastnode)++(180:22pt) edge (\tikzlastnode)
}]
\tikzset{
	bg/.default={},
	bg/.style={execute at end picture={
			\begin{scope}[on background layer]
				\node[xshift=-1mm, yshift=-1mm] (sw) at (current bounding box.south west) {};
				\node[xshift=1mm, yshift=1mm] (ne) at (current bounding box.north east) {};
				\node[xshift=1mm, yshift=-1mm] (nw) at (current bounding box.north west) {};
				\fill[fill=black!10,rounded corners] (sw) rectangle (ne);
				
				\ifx&#1&\else
				\node[anchor=north east, xshift=2pt] at (nw) {#1};
				\fi
			\end{scope}
	}},
}

\newcommand{\Func}[1]{{\mathsf{#1}}}
\newcommand{\Val}{\Func{Val}}
\newcommand{\Inf}{\Func{Inf}}
\newcommand{\Sup}{\Func{Sup}}
\newcommand{\DSum}{\Func{DSum}}
\newcommand{\LimInf}{\Func{LimInf}}
\newcommand{\LimSup}{\Func{LimSup}}
\newcommand{\LimInfAvg}{\Func{LimInfAvg}}
\newcommand{\LimSupAvg}{\Func{LimSupAvg}}

\newcommand{\prefixeq}{\preceq}
\newcommand{\prefix}{\prec}

\newcommand{\suchthat}{\;\ifnum\currentgrouptype=16 \middle\fi|\;}
\let\st\suchthat

\newcommand{\safe}[1]{{\it SafetyCl}(#1)}

\newcommand{\CompClass}[1]{{\textsc{#1}}\xspace}
\newcommand{\PTime}{\CompClass{PTime}}

\newcommand{\PSpace}{\CompClass{PSpace}}
\newcommand{\PSpaceC}{\CompClass{PSpace}-complete\xspace}

\newcommand{\ExpSpace}{\CompClass{ExpSpace}}

\title{QuAK: Quantitative Automata Kit\thanks{This work was supported in part by the ERC-2020-AdG 101020093.}} 
\titlerunning{QuAK: Quantitative Automata Kit}

\author{
	Marek~Chalupa\inst{1}\orcidID{0000-0003-1132-5516} \and 
	Thomas~A.~Henzinger\inst{1}\orcidID{0000-0002-2985-7724} \and 
	Nicolas~Mazzocchi\inst{2,1}\orcidID{0000-0001-6425-5369}\thanks{Corresponding author. N.~Mazzocchi was affiliated with ISTA when his collaboration started.} \and 
	N.~Ege~Sara\c{c}\inst{1}\orcidID{0009-0000-2866-8078}\thanks{Corresponding author.}
}
\authorrunning{M.~Chalupa \and T.~A.~Henzinger \and N.~Mazzocchi \and N.~E.~Sara\c{c}} %

\institute{
	Institute of Science and Technology Austria (ISTA), Austria\\
	\email{\{mchalupa,tah,esarac\}@ista.ac.at}
	\and
	Slovak University of Technology in Bratislava, Slovak Republic\\
	\email{nicolas.mazzocchi@stuba.sk}
}

\date{}

\makeatletter 
\def\orcidID#1{\smash{\href{http://orcid.org/#1}{\protect\raisebox{-1.25pt}{\protect\includegraphics{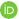}}}}}    
\makeatother

\begin{document}
	\maketitle
	
%	\rednote[inline]{You can leave notes using the command \textbackslash yourInitials\{note\}. \\ \mc{Test.} \tah{Test.} \nm{Test.} \nes{Test.}}

	\begin{abstract}
	System behaviors are traditionally evaluated through binary classifications of correctness, which do not suffice for properties involving quantitative aspects of systems and executions.
	Quantitative automata offer a more nuanced approach, mapping each execution to a real number by incorporating weighted transitions and value functions generalizing acceptance conditions.
	In this paper, we introduce QuAK, the first tool designed to automate the analysis of quantitative automata.
	QuAK currently supports a variety of quantitative automaton types, including $\Inf$, $\Sup$, $\LimInf$, $\LimSup$, $\LimInfAvg$, and $\LimSupAvg$ automata, and implements decision procedures for problems such as emptiness, universality, inclusion, equivalence, as well as for checking whether an automaton is safe, live, or constant.
	Additionally, QuAK is able to compute extremal values when possible, construct safety-liveness decompositions, and monitor system behaviors.
	We demonstrate the effectiveness of QuAK through experiments focusing on the inclusion, constant-function check, and monitoring problems.
	
\end{abstract}

\section{Introduction} \label{sec:introduction}

System behaviors are traditionally seen as sequences of system events, and specifications typically categorize them as correct or incorrect without providing more detailed information.
This binary perspective has long been the cornerstone of formal verification.
However, many interesting system properties require moving beyond this view to systematically reason about timing constraints, uncertainty, resource consumption, robustness, and more, which necessitates a more nuanced approach to the specification, modeling, and analysis of computer systems.

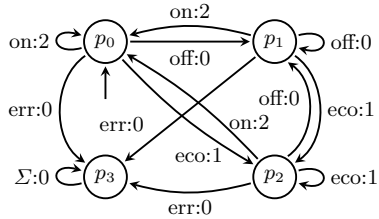
\begin{figure}[t]
	\centering
	\noindent
	\scalebox{0.9}{
		\begin{tikzpicture}[node distance =2cm]
			\node[state, label=center:$p_0$] (0) {};
			\node[yshift=-1cm] (i) at (0) {};
			\node[state, xshift=.5cm, right of = 0, label=center:$p_1$] (1) {};
			\node[state, below of = 1, label=center:$p_2$] (2) {};
			\node[state, below of = 0, label=center:$p_3$] (3) {};
			
			\path[transition]
			(i) edge (0)
			(0) edge[loop left] node[left] {\small on:2} (0)
			(1) edge[bend right=15] node[above] {\small on:2} (0)
			(2) edge[bend right=8, pos=.3] node[right] {\small on:2} (0)
			(0) edge[bend right=0] node[below] {\small off:0} (1)
			(1) edge[loop right] node[right] {\small off:0} (1)
			(2.60) edge[bend right=55, pos=.6] node[left] {\small off:0} (1.300)
			(0) edge[bend right=8,pos=.85] node[yshift=-2pt, left] {\small eco:1} (2)
			(1) edge[bend left=55] node[right] {\small eco:1} (2)
			(2) edge[loop right] node[right] {\small eco:1} (2)
			(0) edge[bend right=55] node[left] {\small err:0} (3)
			(1) edge[bend left=0, pos=.8] node[above left] {\small err:0} (3)
			(2) edge[bend left=15] node[below] {\small err:0} (3)
			(3) edge[loop left] node[left] {\small $\Sigma$:0} (3)
			;	
		\end{tikzpicture}
	}
	\caption{\label{fig:intro} A quantitative automaton $\A$ modeling the power consumption of a device. Associating with $\A$ different value functions, we can specify different aspects of its power consumption, e.g., considering $\LimInfAvg$, the automaton specifies the long-term average power consumption.}
\end{figure}

Quantitative automata~\cite{DBLP:journals/tocl/ChatterjeeDH10} extend standard boolean $\omega$-automata with weighted transitions and a value function that accumulates an infinite sequence of weights into a single value, which generalizes the notion of acceptance condition.
The common value functions include $\Inf$, $\Sup$, $\LimInf$, and $\LimSup$ (respectively generalizing safety, reachability, co-Büchi and Büchi acceptance conditions), as well as $\DSum$ (discounted sum), $\LimInfAvg$ and $\LimSupAvg$ (limit average a.k.a. mean payoff).
Let us consider the quantitative automaton $\A$ given in \cref{fig:intro}, which models the power consumption of a device.
With the value function $\Inf$, it maps each execution to its minimal power consumption, whereas with $\LimInfAvg$ or $\LimSupAvg$ to its long-term average power consumption.
For example, the infinite execution $(\text{off} \cdot \text{on})^\omega$ is mapped to 0 with the value function $\Inf$, and to 1 with $\LimInfAvg$ or $\LimSupAvg$.

The basic decision problems for boolean automata extend to this model naturally.
A quantitative automaton $\A$ is non-empty (resp. universal) with respect to a rational number $v$ iff $\A$ maps some (resp. every) infinite word $w$ to a value at least $v$~\cite{DBLP:journals/tocl/ChatterjeeDH10}.
Notice that problem of non-emptiness (resp. universality) is closely related with computing the maximal (resp. minimal) value achievable by the automaton.
As an example, consider the quantitative automaton given in \cref{fig:intro} with the value function $\LimInfAvg$ and the threshold value $v = 1$.
Evidently, the maximal value achievable by $\A$ in \cref{fig:intro}---its so-called top value---is 2, as witnessed by the word $\text{on}^\omega$, and thus $\A$ is not empty with respect to $v = 1$.
Moreover, an automaton $\A$ is included in (resp. equivalent to) another automaton $\B$ iff, for every infinite word $w$, the value mapped to $w$ by $\A$ is at most (resp. exactly) the value mapped to $w$ by $\B$~\cite{DBLP:journals/tocl/ChatterjeeDH10}.
Beyond their close relation to game theory, these problems are relevant and interesting because solving them allows us to reason about quantitative aspects of systems and their executions.

Introduced around fifteen years ago, the model of quantitative automata has been a significant focus of research~\cite{DBLP:journals/corr/BokerH14,DBLP:conf/lics/BokerHO15,DBLP:conf/concur/MichaliszynO19,DBLP:conf/ijcai/MichaliszynO21,DBLP:conf/lics/Boker24}, with its value functions such as discounted sum and mean payoff being extensively explored in games with quantitative objectives for an even longer period~\cite{shapley1953stochastic,ehrenfeucht1979positional}.
Nonetheless, the notions of monitorability, safety, and liveness for quantitative properties have been introduced and studied only recently~\cite{DBLP:conf/lics/HenzingerS21,DBLP:conf/rv/HenzingerMS22,DBLP:conf/fossacs/HenzingerMS23,DBLP:conf/concur/BokerHMS23}.
These notions and the corresponding problems are important both from a theoretical and a practical perspective, because the study of these problems deepens our understanding of the models of quantitative automata and games, and their solutions can support our verification efforts.
As for boolean properties~\cite{DBLP:journals/dc/AlpernS87,DBLP:journals/fmsd/KupfermanV01}, the ability to check if an automaton is safe lead to specialized verification techniques, e.g., the characterization of safety as continuity may enable verification techniques in domains outside of automata theory~\cite{DBLP:conf/concur/BokerHMS23}.
Despite the potential implications of these results on the practical verification of quantitative properties, no general tool exists for the analysis of quantitative automata.

In this paper, we introduce QuAK, the first general tool designed to automate the analysis of quantitative automata.
QuAK supports a range of quantitative automaton types, including $\Inf$, $\Sup$, $\LimInf$, $\LimSup$, $\LimInfAvg$, and $\LimSupAvg$ automata.
It implements decision procedures for fundamental problems such as emptiness, universality, inclusion, equivalence, constant-check, safety, and liveness.
In particular, it leverages the antichain-based inclusion algorithm for B\"uchi automata~\cite{DBLP:conf/cav/DoveriGM22} by extending the tool FORKLIFT~\cite{forklift}.
Additionally, QuAK provides capabilities for computing top and bottom values of quantitative automata, performing safety-liveness decompositions, and monitoring system behaviors.
QuAK is designed to be free of dependencies beyond the C++ standard library and features an intuitive interface, making it accessible and easy to use.

Modeling beyond-boolean aspects of systems has been considered in several different ways.
One approach considers multi-valued truth domains instead of binary domains~\cite{DBLP:conf/cav/BrunsG99,DBLP:conf/cav/ChechikGD02}.
Another prominent approach involves weighted automata~\cite{DBLP:journals/iandc/Schutzenberger61b}, which extend classical automata by assigning each transition a numerical weight from a semiring whose operations describe how the weights are accumulated.
Tools such as Vaucanson~\cite{DBLP:conf/wia/LombardyPRS03}, Vcsn~\cite{DBLP:conf/wia/DemailleDLS13}, and Awali~\cite{Awali2.3} provide support for the analysis of weighted automata.
The well-established techniques for weighted automata on finite words do not adapt well to the $\omega$-valuation monoid framework necessary for infinite words.
Quantitative automata provide a more intuitive alternative, as they are designed to generalize boolean finite-state $\omega$-automata.
See~\cite{Bok21} for more on the distinction between weighted and quantitative automata.
Another significant approach considers the interaction of digital computational processes with analog physical processes, modeled using automata~\cite{DBLP:journals/tcs/AlurD94,DBLP:conf/lics/Henzinger96} as well as temporal logics~\cite{DBLP:journals/iandc/AlurH93} and implemented in tools such as UPPAAL~\cite{DBLP:journals/sttt/LarsenPY97} and HyTech~\cite{DBLP:conf/hybrid/HenzingerH94a}.
Signal temporal logic~\cite{DBLP:conf/formats/MalerN04}, in particular, has quantitative semantics that allows for reasoning about the degree to which a specification is satisfied or violated, and is implemented in tools such as Breach~\cite{DBLP:conf/cav/Donze10}, S-TaLiRo~\cite{DBLP:conf/tacas/AnnpureddyLFS11}, and RTAMT~\cite{DBLP:conf/atva/Nickovic020}.
Finally, probabilistic verification deals with systems that have inherent uncertainties, such as random failures or probabilistic decision making.
PRISM~\cite{DBLP:conf/cpe/KwiatkowskaNP02} is a widely used tool that allows for the analysis of probabilistic models like Markov chains and Markov decision processes.

%The remainder of this paper is structured as follows:
%Section 2 recalls quantitative automata, their basic decision problems, the notions of quantitative safety and liveness as well as the related constructions and decision procedures (which are implemented in QuAK).
%Section 3 presents the tool QuAK and its key features, where we take a closer look at the implementation details of a selection of algorithms.
%Section 4 shows a brief experimental evaluation of QuAK focusing on the inclusion and monitoring problems.
%Finally, Section 6 concludes with a summary of contributions and directions for future work.
	\section{Quantitative Properties and Automata} \label{sec:preliminaries}

Let $\Sigma = \{a,b,\ldots\}$ be a finite alphabet of letters.
An infinite (resp. finite) word (trace) is an infinite (resp. finite) sequence of letters $w \in \Sigma^\omega$ (resp. $u \in \Sigma^*$).
Moreover, we denote by $\Sigma^+$ the set of non-empty finite words.
Given $u \in \Sigma^*$ and $w \in \Sigma^* \cup \Sigma^\omega$, we write $u \prefix w$ (resp.\ $u \prefixeq w$) when $u$ is a strict (resp.\ nonstrict) prefix of~$w$.
We denote by $|w|$ the length of $w \in \Sigma^* \cup \Sigma^\omega$ and, given $a \in \Sigma$, by $|w|_a$ the number of occurrences of $a$ in $w$.
For $w \in \Sigma^* \cup \Sigma^\omega$ and $0 \leq i < |w|$, we denote by $w[i]$ the $i$th letter of $w$.
An infinite word is \emph{ultimately periodic} (a.k.a, a \emph{lasso word}) iff it is of the form $uv^\omega$ for some $u,v \in \Sigma^*$ with $|v| > 0$.

A \emph{value domain} $\DD$ is a poset.
We assume that $\DD$ is a nontrivial (i.e., $\bot \neq \top$) complete lattice.
Whenever appropriate, we write $0$ or $-\infty$ instead of $\bot$ for the least element $\inf \DD$, and $1$ or $\infty$ instead of $\top$ for the greatest element $\sup \DD$.

A \emph{quantitative property} is a total function $\varPhi : \Sigma^\omega \to \DD$ from the set of infinite words to a value domain.
A boolean property $P \subseteq \Sigma^\omega$ is a set of infinite words.

A \emph{nondeterministic quantitative automaton} on words \cite{DBLP:journals/tocl/ChatterjeeDH10} is a tuple $\A=(\Sigma,Q,\iota,\delta)$, where $\Sigma$ is an alphabet; $Q$ is a finite nonempty set of states; $\iota\in Q$ is the initial state; and $\delta\colon Q\times \Sigma \to 2^{\QQ \times Q}$ is a finite transition function over weight-state pairs.
A \emph{transition} is a tuple $(q,\sigma,x,q')\in Q \times \Sigma \times \QQ \times Q$ such that $(x,q')\in\delta(q,\sigma)$, also written $\trans{q}{\sigma:x}{q'}$.
%\bluenote{in general think are cleaner when the weight is not part of the transition relation} 
We write $\gamma(t)=x$ for the weight of a transition $t=(q,\sigma,x,q')$.
An automaton $\A$ is deterministic if, for all $q\in Q$ and $a\in \Sigma$, the set $\delta(q,a)$ is a singleton.
We require the automaton $\A$ to be \emph{total} (a.k.a. \emph{complete}), meaning that for every state $q\in Q$ and letter $\sigma\in\Sigma$, there is at least one state $q'$ and a transition $\trans{q}{\sigma:x}{q'}$.
For a state $q\in Q$, we denote by $\A^q$ the automaton derived from $\A$ by setting its initial state $\iota$ to $q$.

A run of $\A$ on a word $w$ is a sequence $\rho = \trans{q_0}{w[0]:x_0}{q_1}\trans{}{w[1]:x_1}{q_2}\ldots$ of transitions where $q_0=\iota$ and $(x_i,q_{i+1})\in \delta(q_i,w[i])$.
When $w$ is a finite word, we also write $\rho = \trans{q_0}{w}_{\A}{q'}$ as shorthand, where $q'$ one of the states $\A$ reaches after reading $w$.
For $0 \leq i < |w|$, we denote the $i$th transition in $\rho$ by $\rho[i]$, and the finite prefix of $\rho$ up to and including the $i$th transition by $\rho[..i]$.
Since each transition $t_i$ carries a weight $\gamma(t_i)\in\QQ$, the sequence $\rho$ provides a weight sequence $\gamma(\rho) = \gamma(t_0) \gamma(t_1) \ldots$.
A $\Val$-automaton is equipped with a value function $\Val:\QQ^\omega \to \RR$, which assigns real values to runs of $\A$.
%We assume that $\Val$ is bounded for every finite set of rationals, i.e., for every finite $V \subset \QQ$ there exist $m,M \in \RR$ such that $m \leq \Val(x) \leq M$ for every $x \in V^\omega$.
%Note that the finite set $V$ corresponds to the transition weights of a quantitative automaton, and the specific value functions we consider satisfy this assumption.
The value of a run $\rho$ is $\Val(\gamma(\rho))$. 
The value of a $\Val$-automaton $\A$ on a word $w$, denoted $\A(w)$, is the supremum of $\Val(\rho)$ over all runs $\rho$ of $\A$ on $w$.
The \emph{top value} of a $\Val$-automaton $\A$, denoted $\top_{\A}$ (or $\top$ when $\A$ is clear from the context), is the supremum of $\A(w)$ over all words $w$.
Similarly, the \emph{bottom value} of $\A$, denoted  $\bot_{\A}$ (or $\bot$), is the infimum of $\A(w)$ over all words $w$.

%Note that when we speak of the top value of an automaton or a property expressed by an automaton, we always match its value domain to have the same top value.
%	\textcolor{red}{In particular, given an automaton $\A$, we assume that it defines a property over the domain $\DD = [-\infty, \top_{\A}]$}.
%	The \emph{top value} of a $\Val$-automaton $\A$ is the top value of the property it expresses which we denote by $\top_{\A}$, or simply $\top$ when $\A$ is clear from the context.
%	Note that when we speak of the top value of a property or an automaton, we always match its value domain to have the same top value.

%Two automata $\A$ and $\A'$ are \emph{equivalent}, if they express the same function from words to reals.
%The size of an automaton consists of the maximum among the size of its alphabet, state-space, and transition-space, where weights are represented in binary.

We list below the common value functions for quantitative automata, defined over infinite sequences $v_0 v_1 \ldots$ of rational weights.

\begin{multicols}{2}
	\begin{itemize}
		\item $\displaystyle \Inf(v) = \inf\{v_n \st n \geq 0\}$
		\item $\displaystyle \Sup(v) = \sup\{v_n \st n \geq 0\}$
	\end{itemize}
\end{multicols}

\begin{multicols}{2}
	\begin{itemize}
		\item $\displaystyle \LimInf(v) = \lim_{n\to\infty}\limits\inf\{v_i \st i \geq n\}$
		\item $\displaystyle \LimInfAvg(v) = \LimInf \left(\frac{1}{n} \sum_{i=0}^{n-1} v_i\right)$
		\item $\displaystyle \LimSup(v) = \lim_{n\to\infty}\limits\sup\{v_i \st i \geq n\}$
		\item $\displaystyle \LimSupAvg(v) = \LimSup \left(\frac{1}{n} \sum_{i=0}^{n-1} v_i\right)$
	\end{itemize}
\end{multicols}

\begin{itemize}
	\item For a discount factor $\lambda\in\QQ\cap(0,1)$, $\displaystyle \DSum_{\lambda}(v) = \sum_{i\geq 0} \lambda^i  v_i$
\end{itemize}

All of these classes of automata are $\sup$-closed~\cite[Prop. 2.2]{DBLP:conf/concur/BokerHMS23}: for every finite word $u \in \Sigma^*$ there is a continuation $w \in \Sigma^\omega$ with $\A(uw) = \sup_{w' \in \Sigma^\omega} \A(u w')$.

%When we speak of an automaton (without a specified value function), its value function is one of the above.
%Note that (i) we write $\DSum$ when the discount factor $\lambda\in\QQ\cap(0,1)$ is unspecified, and (ii), $\LimInfAvg$ and $\LimSupAvg$ are also called $\MPL$ and $\MPH$ in the literature.
%Our tool currently supports quantitative automata with the value functions $\Inf$, $\Sup$, $\LimInf$, $\LimSup$, $\LimInfAvg$, and $\LimSupAvg$.

%\begin{example}
%	\TODO
%\end{example}

\subsection{Basic Decision Problems of Quantitative Automata} \label{sec:decision}

\subsubsection{Non-emptiness}
An automaton $\A$ is \emph{non-empty} with respect to a threshold $v \in \QQ$ iff $\A(w) \geq v$ for some $w \in \Sigma^\omega$.
Since the top value of an automaton is achievable by a lasso word~\cite[Thm. 3]{DBLP:journals/tocl/ChatterjeeDH10}, it is easy to see that $\A$ is non-empty with respect to $v$ iff $\top_{\A} \geq v$.
The top value of the common classes of quantitative automata can be computed as follows:
For $\Inf$ and $\Sup$ automata, the top value can be computed by a simple extension of the standard attractor construction; for $\LimInf$ and $\LimSup$ automata, by a simple extension of the standard recurrence construction.
For $\LimInfAvg$ and $\LimSupAvg$ automata, it can be computed by Karp's maximum mean cycle algorithm; for $\DSum$ automata, by solving a game with discounted payoff objectives in graphs with rewards on edges.
All of these computations are in \PTime, therefore the non-emptiness problem is in \PTime for these automata classes.

\subsubsection{Universality}
An automaton $\A$ is \emph{universal} with respect to a threshold $v \in \QQ$ iff $\A(w) \geq v$ for all $w \in \Sigma^\omega$.
For $\Inf$, $\Sup$, $\LimInf$, and $\LimSup$ automata, the universality problem is \PSpaceC~\cite[Thm. 7]{DBLP:journals/tocl/ChatterjeeDH10}.
This is achieved by a simple reduction to the boolean universality problem (of safety, reachability, co-Büchi, and Büchi automata, respectively).
For nondeterministic $\LimInfAvg$ and $\LimSupAvg$ automata, the problem is undecidable~\cite{DBLP:conf/csl/DegorreDGRT10,DBLP:conf/concur/ChatterjeeDEHR10}, and for nondeterministic $\DSum$ automata, it is a long-standing open problem.
Nonetheless, for their deterministic counterparts the problem is in \PTime by a simple product construction~\cite[Thm. 8]{DBLP:journals/tocl/ChatterjeeDH10}.

Notice that an automaton is universal with respect to a threshold $v$ iff its bottom value $\bot$ is at least $v$.
%For a deterministic automaton $\A$, the bottom value can be computed in \PTime as $\bot_{\A} = - \top_{-\A}$ where $-\A$ is a copy of $\A$ in which all the weights are multiplied by $-1$.
For a deterministic automaton $\A$, the bottom value can be computed in \PTime as $\bot_{\A} = - \top_{\hat{\A}}$ where $\hat{\A}$ is a copy of $\A$ in which all the weights are multiplied by $-1$ and the value function is replaced by its dual (e.g., using $\Inf$ instead of $\Sup$).
For a nondeterministic $\Inf$, $\Sup$, $\LimInf$, or $\LimSup$ automaton $\A$, we can compute its bottom value via repeated inclusion checks against a constant automaton:
the largest weight $x$ of $\A$ such that the constant automaton $\B$ with the value $x$ is included in $\A$ equals $\bot_{\A}$, which gives us a \PSpace algorithm.

\subsubsection{Inclusion}
An automaton $\A$ is \emph{included} in another automaton $\B$ iff $\A(w) \leq \B(w)$ for all $w \in \Sigma^\omega$.
Like the universality problem, the inclusion is \PSpaceC for $\Inf$, $\Sup$, $\LimInf$, and $\LimSup$ automata~\cite[Thm. 4]{DBLP:journals/tocl/ChatterjeeDH10}.
As expected, for nondeterministic $\LimInfAvg$ and $\LimSupAvg$ automata, it is again undecidable (by a reduction from universality), and for nondeterministic $\DSum$ automata, it is again open.
When $\B$ is deterministic, the problem is again in \PTime by a simple product construction~\cite[Thm. 5]{DBLP:journals/tocl/ChatterjeeDH10}.
Analogously, $\A$ is \emph{equivalent} to $\B$ iff $\A(w) = \B(w)$ for all $w \in \Sigma^\omega$.
The decidability and complexity results for the inclusion problem carry to equivalence problem.

The standard approach for $\Inf$, $\Sup$, $\LimInf$, and $\LimSup$ automata relies on reducing the problem to boolean inclusion: an automaton $\A$ is included in another automaton $\B$ iff $L(\A, v) \subseteq L(\B, v)$ holds for every weight $v$ of $\A$, where $L(\A, v) = \{w \in \Sigma^\omega \st \A(w) \geq v\}$.
QuAK takes an alternative approach based on antichains~\cite{DBLP:conf/cav/WulfDHR06,DBLP:conf/cav/DoveriGM22}, which we detail in \cref{sec:tool}.

\subsection{Safety and Liveness of Quantitative Automata} \label{sec:safetyliveness}

\subsubsection{Constant-Function Problem}
A quantitative automaton $\A$ defines a constant function iff there exists $c \in \RR$ such that $\A(w) = c$ for all $w \in \Sigma^\omega$.
For all common classes of quantitative automata, deciding whether they define a constant function is \PSpaceC~\cite[Prop. 3.2, Thms. 3.3 and 3.7]{DBLP:conf/concur/BokerHMS23}.
Deciding whether an automaton defines a constant function is closely related with deciding its safety and liveness~\cite{DBLP:conf/concur/BokerHMS23}.
As we will discuss below, this is especially important for limit average automata whose equivalence is undecidable and for discounted sum automata whose universality is open.
For limit average automata, the constant-function problem can be solved by a reduction to the limitedness problem of distance automata.
For discounted sum automata, the problem is reduced to the universality problem of nondeterministic finite automata on finite words.
For the remaining classes of automata, one can simply check if the given automaton is universal with respect to its top value.

\subsubsection{Safety}
The boolean membership problems asks, given a boolean property $P \subseteq \Sigma^\omega$ and a word $w \in \Sigma^\omega$, whether $w$ belongs to $P$.
Then, a boolean property $P$ is safe iff every word $w$ that is not a member of $P$ has a finite prefix $u \prefix w$ such that for every continuation $w'$ the word $u w'$ is not a member of $P$.
%$w \notin P$ has a prefix $u \prefix w$ such that $u w' \notin P$ for all $w' \in \Sigma^\omega$.
The quantitative decision problems discussed above implicitly generalize the boolean membership problem to the quantitative setting by asking, given an automaton $\A$, a value $v$, and a word $w$, whether the value $\A(w)$ is at least $v$.
The quantitative generalization of safety reflects this view: a quantitative property $\varPhi$ is safe iff every wrong membership query has a finite witness for the violation.

\begin{definition}[\cite{DBLP:conf/fossacs/HenzingerMS23}]
	A quantitative property $\varPhi : \Sigma^\omega \to \DD$ is \emph{safe} iff for every $w \in \Sigma^\omega$ and $v \in \DD$ with $\varPhi(w) \not \geq v$, there exists a finite prefix $u \prefix w$ such that $\sup_{w' \in \Sigma^\omega} \varPhi(u w') \not \geq v$.
\end{definition}

The safety closure $\safe{\varPhi}$ of a property $\varPhi$ ensures its safety by minimally increasing the value of each word, and a property is safe iff it is equal to its safety closure (see~\cite[Defn. 5, Prop. 6, Thm. 9]{DBLP:conf/fossacs/HenzingerMS23}).
For the common classes of quantitative automata, we can compute their safety closure in \PTime by assigning the top value of each state to all its incoming transitions~\cite[Thm.~4.18]{DBLP:conf/concur/BokerHMS23}.

The decision procedures for safety in quantitative automata depend on the value functions, with $\Inf$ and $\DSum$ automata always defining safety properties~\cite[Thm. 4.15]{DBLP:conf/concur/BokerHMS23}, while for $\Sup$, $\LimInf$, and $\LimSup$ automata, safety check is \PSpaceC thanks to the decidability of their equivalence~\cite[Thm. 4.22]{DBLP:conf/concur/BokerHMS23}.
For $\LimInfAvg$ and $\LimSupAvg$ automata, despite the undecidability of their equivalence problem, their safety can be decided in \ExpSpace by using their constant-function check as a subroutine~\cite[Thm. 4.23]{DBLP:conf/concur/BokerHMS23}.

\subsubsection{Liveness}

As for boolean safety, the notion of boolean liveness takes the membership problem to its basis.
A boolean property $P$ is live iff for every finite word $u$ (even if $u$ is prefix of some word that is not a member of $P$) there is a continuation $w$ such that the word $uw$ is a member of $P$.
Quantitative liveness extends this membership-based view: a quantitative property $\varPhi$ is live iff, whenever a property value is less than $\top$, there exists a value $v$ for which the wrong membership query $\varPhi(w) \geq v$ can never be dismissed by any finite witness $u \prec w$.

\begin{definition}[\cite{DBLP:conf/fossacs/HenzingerMS23}]
	A property $\varPhi : \Sigma^\omega \to \DD$ is \emph{live} when for all $w \in \Sigma^\omega$, if $\varPhi(w) < \top$, then there exists a value $v \in \DD$ such that $\varPhi(w) \not \geq v$ and for all prefixes $u \prec w$, we have $\sup_{w' \in \Sigma^\omega} \varPhi(uw') \geq v$.
\end{definition}

In the quantitative setting, liveness is characterized by the safety closure operation, where a quantitative property $\varPhi$ is live iff $\varPhi(w) < \safe{\varPhi}(w)$ for all words $w$ with $\varPhi(w) < \top$~\cite[Thm. 37]{DBLP:conf/fossacs/HenzingerMS23}.
For $\sup$-closed quantitative properties (like those defined by quantitative automata thanks to~\cite[Prop. 2.2]{DBLP:conf/concur/BokerHMS23}), a property is live iff its safety closure defines a constant function $\top$~\cite[Thm.~5.7]{DBLP:conf/concur/BokerHMS23}.

To decide liveness for a quantitative automaton $\A$, one can check if $\safe{\A}$ maps all words $w$ to $\top$, which involves checking the universality of $\safe{\A}$ with respect to $\top$.
Note that $\safe{\A}$ is an $\Inf$ automaton when $\Val \in \{\Inf, \Sup, \LimInf, \LimSup, \LimInfAvg, \LimSupAvg\}$, and a $\DSum$ automaton when $\Val = \DSum$.
For $\Inf$ automata, this problem is \PSpaceC.
For $\DSum$ automata, despite their universality is open, checking whether their liveness check is \PSpaceC~\cite[Thm. 5.9]{DBLP:conf/concur/BokerHMS23} since checking whether they define a constant function is \PSpaceC.

\subsubsection{Safety-Liveness Decompositions}

Every boolean property $P$ is the intersection of a boolean safety property $S$ and a boolean liveness property $L$.
In this decomposition, $S$ is the boolean safety closure of $P$, while $L$ is the union of $P$ with the complement of $S$.
In a way, the liveness part $L$ balances the safety closure $S$ by not including the words that belong to the difference of $S$ and $P$.

Every quantitative property $\varPhi$ is the pointwise minimum of its safety closure $\safe{\varPhi}$ and a liveness property $\varPsi$~\cite[Thm. 44]{DBLP:conf/fossacs/HenzingerMS23}.
In particular, the liveness component $\varPsi$ is defined similarly to its boolean analogue: for every word $w$, we have $\varPsi(w) = \top$ if $\varPhi(w) = \safe{\varPhi}(w)$, and $\varPsi(w) = \varPhi(w)$ if $\varPhi(w) < \safe{\varPhi}(w)$.
For deterministic  $\Sup$, $\LimInf$, and $\LimSup$ automata, the decomposition works the same way on the level of transition weights (see \cite[Thms. 5.10 and 5.11]{DBLP:conf/concur/BokerHMS23}), which is achievable in \PTime.
Note that $\Sup$ and $\LimInf$ automata are determinizable~\cite[Thm. 13]{DBLP:journals/tocl/ChatterjeeDH10}.

%For some classes of quantitative automata, the decomposition works the same way on the level of transition weights (see \cite[Thms. 5.10 and 5.11]{DBLP:conf/concur/BokerHMS23}).
%Recall that the safety closure of a quantitative automaton preserves the transition structure.
%Consider a deterministic $\Sup$, $\LimInf$, or $\LimSup$ automaton $\A$ and its safety closure $\safe{\A}$.
%The liveness component $\B$ is a copy of $\A$ where, for each transition $t$, if the weight of $t$ is the same in $\A$ and $\safe{\A}$, it takes the weight $\top_{\A}$ in $\B$, and otherwise, it takes the weight of $t$ in $\A$.
%Overall, this decomposition is achievable in \PTime.
%Note that $\Sup$ and $\LimInf$ automata are determinizable~\cite[Thm. 13]{DBLP:journals/tocl/ChatterjeeDH10}.

	\section{The Tool} \label{sec:tool}

QuAK implements the decision procedures and constructions described in \cref{sec:decision,sec:safetyliveness}.
In particular, let $\A$ and $\B$ be $\Val$ automata where $\Val \in \{\Inf, \Sup, \LimInf, \LimSup, \LimInfAvg, \LimSupAvg\}$, and let $v \in \QQ$ be a rational number.
QuAK currently supports the following operations (whenever known to be computable):
\begin{enumerate}
	\item Check if $\A$ is non-empty with respect to $v$.
	\item Check if $\A$ is universal with respect to $v$.
	\item Check if $\A$ is included in $\B$.
	\item Check if $\A$ defines a constant function.
	\item Check if $\A$ defines a safety property.
	\item Check if $\A$ defines a liveness property.
	\item Compute the top value $\top$ of $\A$.
	\item Compute the bottom value $\bot$ of $\A$.
	\item Compute the safety closure of $\A$.
	\item Compute the safety-liveness decomposition of $\A$.
	\item Construct and execute a monitor for $\A$.
\end{enumerate}
The tool is written in C++ using the standard library, and is available at \url{https://github.com/ista-vamos/QuAK} together with detailed instructions on its usage.
%In this section, we describe the automata representation, our antichain-based inclusion algorithm, and monitoring approach for quantitative properties.
In this section, we describe the automata representation, our antichain-based inclusion algorithm, implementation of the constant-function check for limit average automata, and monitoring approach for quantitative properties. 
The rest follows the descriptions in \cref{sec:decision,sec:safetyliveness}.

\subsubsection{Automata Representation}
We represent automata in a way that makes the algorithms as efficient as possible while keeping their implementation convenient and maintainable.
Here, we explain some choices we made for this sake.

Automata objects do not have a value function because it may be useful to interpret the same transition structure in different ways (like in \cref{fig:intro}).
The user needs to specify the value function when a decision procedure or a construction is called on an automaton.
Moreover, each automaton contains a directed acyclic graph representing its strongly connected components (SCCs), which is constructed once the automaton is created.
Each state has a tag representing the SCC it belongs to.
Moreover, in addition to storing the outgoing transitions of a state, we also store the incoming transitions.
These are useful when computing the top value, constructing the safety closure, and determinizing the safety closure of limit average automata (for deciding their safety).
Finally, while boolean automata has a fixed domain $\{0,1\}$, each quantitative automaton may define a different domain.
To address this, each automaton has two numerical variables representing the minimum and maximum of its domain, whose values are taken as the minimum and maximum of the automaton's weights by default.

\subsubsection{Antichain-based Inclusion Algorithm}\label{ssec:antichain-alg}

The language inclusion of B\"uchi automata (a.k.a. $\LimSup$) is known to be \PSpaceC.
Algorithms that behave well in practice have been investigated for decades and remains an active research field.
Among others, FORKLIFT~\cite{forklift} uses the Ramsey-based technique and leverages the antichains heuristic.
The algorithm to decide whether $L(A) \subseteq L(B)$ holds searches counterexamples and runs membership queries.
The Ramsey-based approach prunes the search for inclusion violation by discarding candidates of $L(A)$ which are subsumed by others words of $L(A)$ with respect to a given well-quasiorder.
Termination comes from the mathematical properties of well-quasiorders guaranteeing that only finitely many candidates will be kept after pruning.
Correctness is trivial: if a kept candidate violates $L(A) \subseteq L(B)$ then the inclusion does not holds.
To guarantee completeness, we require the quasiorder to fulfill, for all candidate $w \in \Sigma^\omega$ subsumed by $w_0\in \Sigma^\omega$, that $w_0\in L(B)$ implies $w\in L(B)$.
Hence, if all candidates belong to $L(B)$ then so do the discarded ones.
The antichains heuristic allows a symbolic fixpoint computation of the remaining candidates~\cite{DBLP:conf/cav/WulfDHR06}.

Language inclusion can be decided by reasoning solely on ultimately periodic words (a.k.a. lasso words).
So, the candidates are words of the form $uv^{\omega}$, where $u\in\Sigma^*$ and $v\in\Sigma^+$ are called a stem and a period, respectively.
\cite{DBLP:conf/concur/DoveriGPR21} provides an algorithm that uses two quasiorders: one for the stems and one for the periods.
Since using different quasiorders yields more pruning when searching for a inclusion violation, \cite{DBLP:conf/cav/DoveriGM22} considers using an unbounded number of quasiorders called \textsf{FORQ}: one for the stems and a family of quasiorders for the periods each of them depending on a distinct stem.
It is worth emphasizing that each quasiorder requires a fixpoint computation, and thus, the more quasiorders are handled, the more the antichains heuristic is leveraged.

The novelty of FORKLIFT lies in the use of \textsf{FORQ} to discard candidates.
Below, we generalize \textsf{FORQ}s for B\"uchi automata defined in~\cite{DBLP:conf/cav/DoveriGM22} to support any $\LimSup$ automata.

\begin{definition}\label{def:forq}
	Let $\B = (\Sigma,Q,\iota,\delta)$ be a $\LimSup$ automaton over the weights $W = \{\gamma(t) \st \text{$t$ is a transition of $\B$}\}$.
	The structural \textsf{FORQ} of $\B$ is the pair $({\precsim^{\B}}, \{{\precsim_{u}^{\B}}\}_{u \in \Sigma^{*}})$ where the quasiorders are defined by:
	\begin{align*}
		u_1 \precsim^{\B} u_2 & \iff \textsf{Tgt}_{\B}(u_1) \subseteq \textsf{Tgt}_{\B}(u_2)
		\\
		v_1 \precsim^{\B}_{u} v_2 &\iff \textsf{Cxt}_{\B}(\textsf{Tgt}_{\B}(u), v_1) \subseteq \textsf{Cxt}_{\B}(\textsf{Tgt}_{\B}(u), v_2)
	\end{align*}
	with $\textsf{Tgt}_{\B} \colon \Sigma^* \rightarrow 2^{Q}$ and $\textsf{Cxt}_{\B}\colon 2^{Q} \times \Sigma^+ \rightarrow 2^{Q \times Q \times W}$ such that
	\begin{align*}
		\textsf{Tgt}_{\B}(u) &= \{q' \in Q \st \iota \xrightarrow{u}_{\B} q' \}
		\\
		\textsf{Cxt}_{\B}(S, v) &= \{ (q, q', x) \st q \in S, \rho = q \xrightarrow{v}_{\B} q', \text{and $x$ is the maximum} \\
		&\hspace{4.5cm}\text{of the weight sequence of $\rho$}\}
	\end{align*}
\end{definition}

The modification appears in the definition of $\textsf{Cxt}_{\B}$ where $x$ ranges over the weights of $\B$ instead of $\{\bot, \top\}$.
Extending all the properties on structural \textsf{FORQ} established by~\cite{DBLP:conf/cav/DoveriGM22} to this definition is straightforward, and implies the soundness of our inclusion algorithm for $\LimSup$ automata.
To use this algorithm for other classes of automata, we translate $\Inf$, $\Sup$, and $\LimInf$ automata to $\LimSup$ automata in \PTime for inclusion queries.

We highlight the remaining technical modifications below.
%\orangenote{Few technical modifications from FORKLIFT to QuAK are commented in the Latex source. Tell me if some more explanations/details are need.}
\begin{itemize}
\item
	Given a stem $u\in\Sigma^*$, the data structure used for the fixpoint computation of $\textsf{Cxt}_{\B}$ carries (as in FORKLIFT) a period $v \in\Sigma^+$, a context $\textsf{Cxt}_{\B}(\textsf{Txt}_{\B}(u), v)$, and (in addition to FORKLIFT) the value of $\A$ over $uv^\omega$.
% \item
% 	In FORKLIFT, the data structure used for the fixpoint computation of $\textsf{Cxt}_{\B}$ stores a triplet $(q, q', k) \in \textsf{Cxt}_{\B}(\textsf{Txt}_{\B}(u), v)$ once when $k=0$ and twice when $k=1$.
% 	QuAK can mimic this behavior, but the redundancy of the information may reach the number of weights of $\B$.
% 	A compilation option provides an implementation for QuAK without this redundancy.
% 	We call this optimization \texttt{no-redundancy}.
\item 
	In FORKLIFT, the fixpoint computation of $\textsf{Cxt}_{\B}$ does not leverage SCCs.
	A compilation option provides an implementation for QuAK that computes $\textsf{Cxt}_{\B}$ while only considering intra-SCC transitions.
%	For $\A$, it is clear that transitions leaving an SCCs cannot bring extra information.
%	For $\B$, however, the soundness is not immediate, and requires the \textsf{FORQ} to satisfy the picky constraint defined in~\cite{DBLP:conf/cav/DoveriGM22}, which our generalization fulfills as well.
	We call this optimization \texttt{scc-search}.
\end{itemize}

\subsubsection{Constant-function Check for Limit Average Automata}\label{ssec:constant}
Checking whether a limit average automaton $\A$ is constant can be done by a reduction to the limitedness problem of distance automata~\cite[Thm. 3.7]{DBLP:conf/concur/BokerHMS23}.
To simplify our implementation, we consider a reduction to the universality problem of $\LimInf$ automata.
In essence, our reduction follows~\cite[Thm. 3.7]{DBLP:conf/concur/BokerHMS23} except for in two points.
First, after removing negative-weighted edges by Johnson's algorithm, instead of constructing a distance automaton, we flip the weights again to construct a limit average automaton (which resolves nondeterminism by $\sup$ and has the same value function as the input automaton).
This yields an automaton $\B$ with non-positive transition weights.
Then, we construct a $\LimInf$ automaton $\C$ by mapping negative weights of $\B$ to 0, and 0-valued weights to 1.
Note that it may not hold that $\B(w) < 0$ iff $\C(w) < 1$ for all words $w$, but since $\C$ recognizes an $\omega$-regular language, we can show that $\B$ is constant iff $\C$ is universal with respect to 1.

More specifically, the reduction goes as follows.
Let $\A$ be a $\LimInfAvg$ (resp. $\LimSupAvg$) automaton.
First, obtain $\A_1$ by subtracting $\top$ from all transition weights of $\A$.
%We have $\A(w) < \top$ iff $\A_1(w) < 0$ for all words $w$ (since $\A_1(w) = \A(w) - \top$ for all $w$).
We have $\A_1(w) = \A(w) - \top$ for all words $w$.
Then, obtain $\A_2$ by multiplying by $-1$ all transition weights of $\A_1$, resolving nondeterminism by $\inf$, and taking the value function $\LimSupAvg$ (resp. $\LimInfAvg$).
%We have $\A_1(w) < 0$ iff $\A_2(w) > 0$ for all words $w$ (since $\A_2(w) = -\A_1(w)$ for all $w$).
We have  $\A_2(w) = -\A_1(w)$ for all words $w$.
Then, obtain $\A_3$ by using Johnson's algorithm to remove transitions with negative weights of $\A_2$.
%We have $\A_2(w) > 0$ iff $\A_3(w) > 0$ for all words $w$ (since $\A_3(w) = \A_2(w)$ for all $w$). % wrong in paranth
We have $\A_2(w) > 0$ iff $\A_3(w) > 0$ for all words $w$.
Then, obtain $\B$ by multiplying by $-1$ all transition weights of $\A_3$, resolving nondeterminism by $\sup$, and taking the value function $\LimInfAvg$ (resp. $\LimSupAvg$).
%We have $\A_3(w) > 0$ iff $\B(w) < 0$ for all words $w$ (since $\B(w) = -\A_3(w)$ for all $w$).
We have $\B(w) = -\A_3(w)$ for all words $w$.
Note that all transitions of $\B$ have non-positive weights.
Finally, obtain $\C$ by updating the weights of $\B$ as follows and taking the value function $\LimInf$:
if a transition has weight 0, then its new weight is 1; otherwise (weight less than 0), then its new weight is 0.
%
%By construction, $\A$ is constant iff $\B$ is constant.
%We argue that $\B$ is constant iff $\C$ is universal (with respect to 1).
%If $\B$ is not constant, it is easy to see that $\C$ is not universal as the same word witnesses both.
%Suppose $\C$ is not universal.
%Notice that $\C$ is essentially a co-Büchi automaton that defines a non-empty $\omega$-regular language $L$.
%As the complement $L$ is also regular, there is an ultimately periodic word $w$ such that $\C(w) = 0$ and this value is achieved by a lasso run in which the weight 0 appears in the infinitely repeated part.
%Since $\B$ differs from $\C$ only in transition weights, $\B$ also has a lasso run over $w$, where some negative weight appears in the infinitely repeated part. 
%???
%%we have $\A_4(w) < 0$ iff all runs of $\A_4$ on $w$ yield a weight sequence with infinitely many negative values iff all runs of $\A_5$ on $w$ yield a weight sequence with infinitely many 0s iff $\A_5(w) < 1$. % WRONG
%Therefore, $\A$ defines a constant function iff $\C$ is universal with respect to 1.
%Note that we can directly construct $\C$ from $\A$ in \PTime.
%

%%%%%%%%%%%%%%%%%%%%%%%%%%%%%%%%%%%%%%%%%%%%%
% nicolas begin
%%%%%%%%%%%%%%%%%%%%%%%%%%%%%%%%%%%%%%%%%%%%%%
By construction, $\A$ is constant iff $\B$ is constant.
We argue that $\B$ is constant iff $\C$ is universal (with respect to 1).
If there is a word $u\in\Sigma^\omega$ such that $\B(u)<0$, then all runs of $\B$ over $u$ visit infinitely often some negative weight.
Thus, $\C(u)<1$ comes as a direct consequence of this implication.
Note, however, that the reciprocal is not true, i.e., all runs of a word could visit infinitely often some negative weight while being mapped to $0$ by $\B$.
Now, if there is a word $v\in\Sigma^\omega$ such that $\C(v)<1$, then there exists also an ultimately periodic word $w\in\Sigma^\omega$ such that $\C(w)<1$.
This is because $\C$ is a co-B\"uchi automaton that defines a non-empty $\omega$-regular language.
Let $w$ be of the form $w_1^{}w_2^{\omega}$. 
We define $x=|w_1|$, $y=|w_2|$, and let $n$ be the number of states of $\C$.
Suppose towards contradiction that some run of $\C$ over $w$ visits only the weight $1$ for $x+yn$ consecutive transitions.
It implies that this run visits twice the same state at the end of the period $w_2$ while visiting only the weight $1$ in between, which exhibits another run of $\C$ over $w$ of value $1$, and thus leads to the contradiction $\C(w)=1$.
Hence, all runs of $\C$ over $w$ periodically visit the weight $0$ after $x+yn$ transitions.
Since $\B$ differs from $\C$ only in transition weights, all runs of $\B$ over $w$ periodically visit some negative weight after $x+yn$ transitions, therefore $\B(w)<0$.
In conclusion, $\A$ defines a constant function iff $\C$ is universal (with respect to 1).
Note that we can directly construct $\C$ from $\A$ in \PTime.
%%%%%%%%%%%%%%%%%%%%%%%%%%%%%%%%%%%%%%%%%%%%%
% nicolas end
%%%%%%%%%%%%%%%%%%%%%%%%%%%%%%%%%%%%%%%%%%%%%%

\subsubsection{Monitoring}
Given a specification represented as a deterministic quantitative automaton $\A$, QuAK is able create a monitor object that stores an array of top values (storing the top value of $\A^q$ for each state $q$ of $\A$), an array of bottom values (storing the bottom value of $\A^q$ for each state $q$ of $\A$), and a pointer to the current state of $\A$ (initialized as the initial state of $\A$).
A monitor object can read input letters incrementally while getting the next state $q$ of $\A$ and maintaining the lowest and highest values achievable from $q$, namely, the bottom and top values of $\A^q$.
In addition, we implement running average monitors for limit average automata.

	\section{Experimental Evaluation} \label{sec:experiments}

%We conducted a set of experiments to evaluate the performance and applicability of QuAK.
We evaluated QuAK in a set of experiments.
In particular, we measure the performance
our antichain-based inclusion algorithm and compare it to the standard
algorithm based on repeated reduction to language inclusion of Büchi automata.
These experiments include also the measurement of the impact of the \texttt{scc-search} optimization described in \cref{sec:tool}.
Next, we evaluate the runtime of checking if an automaton defines a constant funcion.
Finally, we use QuAK to runtime monitor the smoothness of a controller
for a drone to show that the tool can be used in the context of quantitative runtime monitoring.
The artifact to reproduce the experiments can be found at \url{https://doi.org/10.5281/zenodo.13132069}.

\paragraph{Setup}
QuAK was compiled with -O3 and link-time optimizations enabled.
The \texttt{scc-search} optimization was enabled for all experiments except a part of those that aimed at evaluating
this optimization (\cref{ssec:scc-search}).
All experiments ran on machines with \emph{AMD EPYC} CPU with the frequency 3.1\,GHz. 
The time limit was set to 100\,s wall time.

\paragraph{Benchmarks}
Because of the lack of benchmarks for quantitative automata,
we used randomly generated quantitative automata.
All automata are complete (i.e.,~every state has an outgoing transition for each symbol in the alphabet) and have weights between -10 and 10
chosen uniformly at random.
An automaton that has $n$ states can have up to $n|\Sigma| + 2n + 1$ edges where $\Sigma$ is the alphabet.
As a result, the generated automata are non-deterministic.
The number of states and the size of alphabet differ in the experiments and are always explicitely mentioned.

\subsection{Comparing Inclusion Algorithms}

\newcommand{\antichains}{\emph{Antichains}\xspace}
\newcommand{\standard}{\emph{Standard}\xspace}

In this subsection, we compare the standard approach to compute the
quantitative automata inclusion (referred to as \standard)
with our antichain-based inclusion algorithm (referred to as \antichains).
The implementation of the standard approach 
uses the boolean version of \textsc{FORKLIFT} to decide the inclusion of boolean
automata.
Both algorithms are implemented in QuAK.
%For these experiments, both the optimization \texttt{scc-search} based on SCC analysis and the optimization \texttt{no-redundancy} based on context reduction were enabled.

\begin{figure}[t]
\includegraphics[width=.48\textwidth]{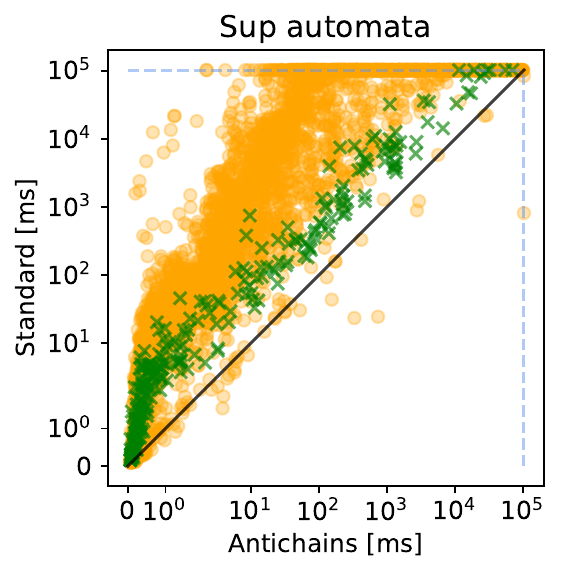}
\includegraphics[width=.48\textwidth]{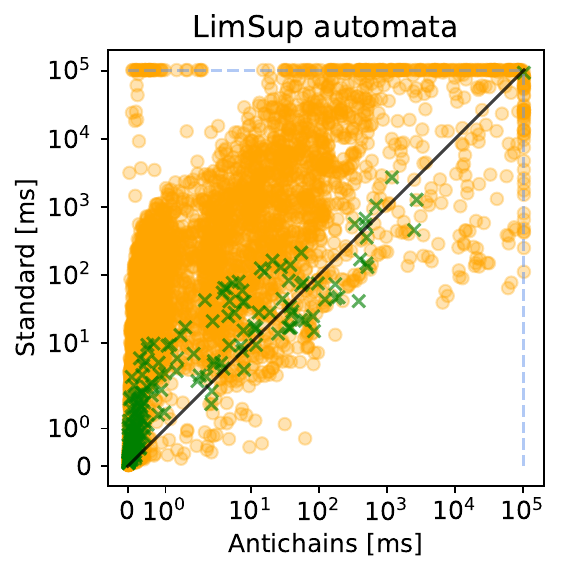}
  \caption{CPU time in milliseconds of running \antichains ($x$~axis)
           and \standard ($y$~axis) inclusion algorithms on random automata with
           2--32 states where at least one algorithm finished within time limit.
           The alphabet has 2.
           Orange dots are for pairs of automata that are not included and
           green crosses are for included automata.
           The scatter plot on the left is for $\Sup$ and on the right
           for $\LimSup$ value function.
           }
  \label{fig:cputime_inclusion}
\end{figure}

\cref{fig:cputime_inclusion} shows the CPU time
of running \standard and \antichains algorithms for $\Val \in \{\Sup, \LimSup\}$.
We used 100 random automata with 2--32 states and with 2-symbol alphabet.
Algorithms were ran for each possible pair of the automata,
which results in 10000 inclusion checks.
In the plots, we show only the runs where at least one algorithm decided the inclusion.

The algorithm \antichains is almost always faster, often significantly, and it can finish
in a lot of cases when \standard reaches the time limit (points on the blue dashed line).
The \standard algorithm internally runs (the boolean version of) \antichains algorithm multiple times for each weight,
and therefore it is expected that \antichains should be faster most of the times.

\subsubsection{Evaluating Optimizations of Inclusion Algorithms}
\label{ssec:scc-search}

\begin{figure}[t]
\includegraphics[width=.48\textwidth]{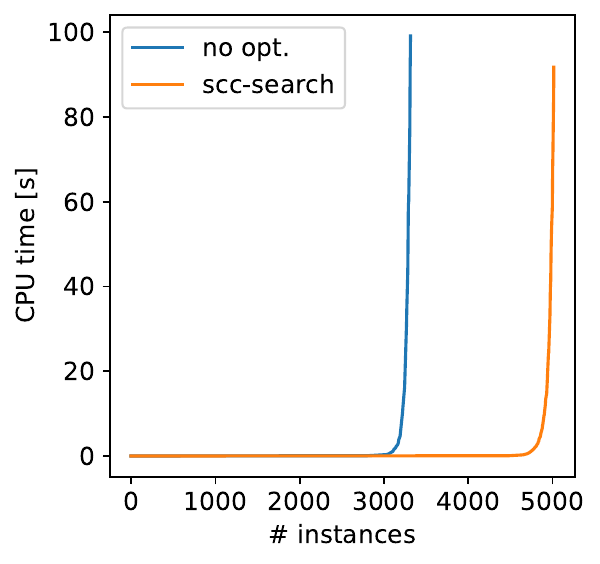}
\includegraphics[width=.48\textwidth]{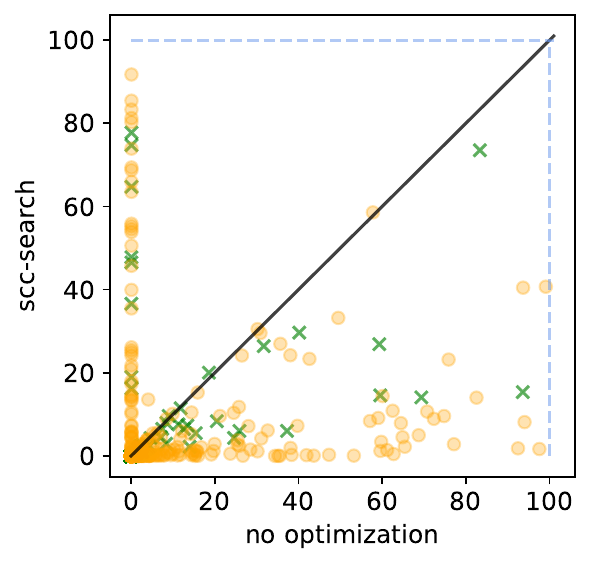}
  \caption{CPU time of running \antichains algorithm with and without the
           \texttt{scc-search} optimization on the benchmarks from
           \cref{fig:cputime_inclusion}.
           In the left plot, the $x$ axis shows how many instances the inclusion algorithms are
           able to decide given the time limit on the $y$ axis.
           The right plot compares the runtime per instance.
           There, orange dots are for pairs of automata that are not included and
           green crosses are for included automata.
           The plots are for $\Sup$ automata and time is in seconds.
           }
  \label{fig:cputime_opts}
\end{figure}

The results in \cref{fig:cputime_inclusion} are for QuAK that is
compiled with the \texttt{scc-search} optimization (see \cref{ssec:antichain-alg}).
Plots in \cref{fig:cputime_opts} show that this optimization
significantly improves the runtime.
The plot on the left shows how many instances
(the $x$ axis) can be decided given the time limit is set to the value on the $y$ axis.
The optimization allows to decide nearly 2000 more instances in under 2 seconds.
The plot on the right shows that the optimization also hurts in some cases.
Nevertheless, it helps with approximately 90\% of the considered automata.

\subsection{Evaluating Constant-function Checking}

\begin{figure}
\includegraphics[width=\textwidth]{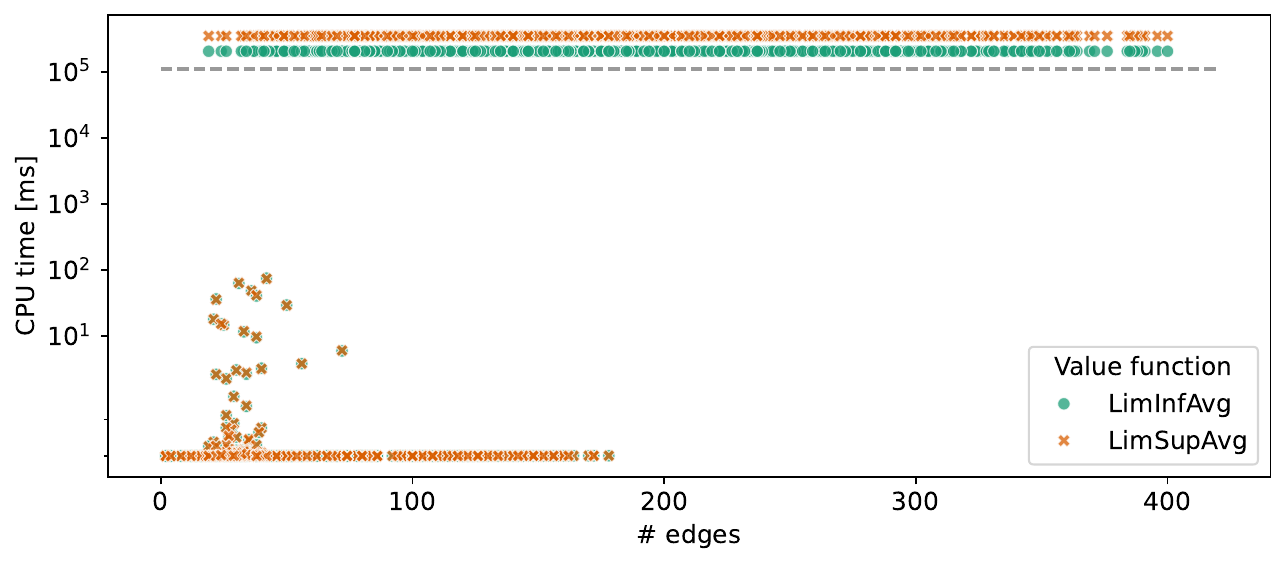}
  \caption{CPU time of deciding if an automaton defines a constant function. 
           Points representing timeouts were moved above the timeout line (the dashed line)
           and separated (with no particular order with respect to the vertical axis).
           }
  \label{fig:constant_fun}
\end{figure}

To evaluate the constant-function checking algorithm for limit-average automata, we generated 1000 random automata with a 4-symbol alphabet and 1--100 states.
The results of running the algorithm on these automata are summarized in \cref{fig:constant_fun}.

The computational complexity of the algorithm increases steeply: for larger automata, the result is typically computed either quickly or not at all.
While the algorithm times out on many instances, the results suggest that deciding whether an automaton is constant remains feasible in certain cases.
Notably, all instances that do not time out in our experiments fall into one of two categories:
they are either deterministic, for which the problem is in \PTime, or the resulting co-Büchi automaton has a very low density of accepting edges, for which the antichain-based algorithm finds witnesses for non-universality more easily.

%To evaluate the constant-function checking algorithm for limit average automata, we generated 1000 random automata with a 4-symbol alphabet and 1--100 states.
%The results of running the algorithm on these automata are summarized in \cref{fig:constant_fun}.
%The plot suggests that deciding if an automaton is constant is feasible for many $\LimInfAvg$ and $\LimSupAvg$ automata.
%Nonetheless, we can see that the computational complexity grows very steeply (after all, the problem is PSPACE-complete): if the automaton is not very small, then the result is either computed very quickly, or not at all.
%% Because there are quickly solved instances for any number of edges, the runtime of the algorithm must depend more on the structure of the automata than on its size. -- NOT TRUE UNLESS THE AUTOMATA ARE TRIMMED

%This is caused by the fact that for these value functions, we use the reduction to universality of $\LimInf$ automata
%(see \cref{ssec:constant}), while for other value functions the problem is solved by checking whether the top value and the bottom value are equal, where the bottom value is computed by a repeated universality check for different weights.
%But also for limit average value functions, we can see that the computational complexity grows very steeply (after all, the problem is PSPACE-complete): if the automaton is not very small, then the result is either computed very quickly, or not at all.

\subsection{Runtime Monitoring}

\begin{figure}[t]
\begin{center}

\begin{minipage}[t]{0.5\textwidth}
  \vspace{0pt}
  \includegraphics[width=\textwidth]{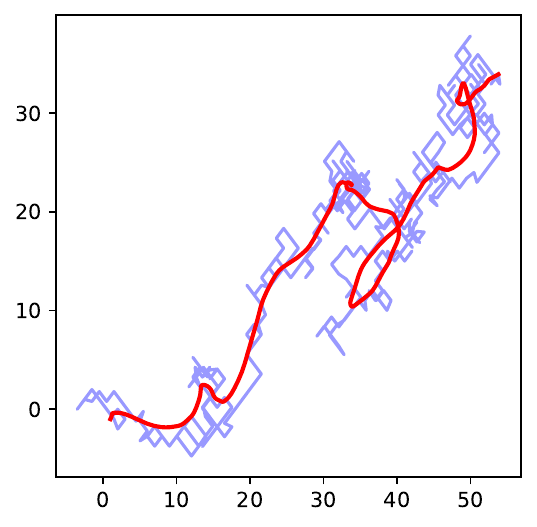}
\end{minipage}%
\begin{minipage}[t]{0.5\textwidth}
    \setlength{\tabcolsep}{.4em}
    \vspace{5em}
    \hspace{1em}
\begin{tabular} {l r r r r}
    \toprule
    &\multicolumn{2}{c}{Original} & \multicolumn{2}{c}{Smoothed}\\
    \cmidrule(r){2-3}
    \cmidrule(l){4-5}
    \em Controller & \em EC & \em Score & EC & \em Score \\
    \midrule
    sd=1 & 9.95 & 12.83 & 3.40 & 0.15 \\
    sd=5 & 12.42 & 15.97& 2.42 & 0.82 \\
    \bottomrule
  \end{tabular}
\end{minipage}
\end{center}

\caption{Results of monitoring the smoothness of a drone controller on an
  erratic and smoothed trajectory (\emph{Original} and \emph{Smoothed}, resp.).
  The situation is depicted on the left where the erratic trajectory is blue and the smoothed
  one is red. Only a part of the trajectories is shown.
  In the table on the right,
  \emph{Score} is the value computed by the monitor and \emph{EC} is the energy
  consumption of the drone on the trajectory.
  The lower is the score, the smoother should be the trajectory.
  All numbers are averages from 3 simulations.}
    \label{tab:drone}
\end{figure}

We experimented with using quantitative automata for runtime monitoring.
Our use case is to monitor the smoothness of controllers of cyber-physical systems (CPS)~\cite{MysoreM0S21},
which means that the actions issued by a CPS controller should always cause only
a relatively small change in the state of the CPS.
For example, a controller of a drone should not instruct it to 
immediately change to the opposite of the current direction.
Controllers that are not smooth can lead to increased energy consumption
or even hardware failures~\cite{MysoreM0S21}.
%\todo{cite}
%(amortization of moving parts, changing controllers on runtime)

We monitored a flying drone in a simulated environment.
For simplicity, we assumed that the drone is a point with mass 1
and its energy consumption is equal to the sum of forces generated by the thrust of its engines.
Each action issued by a controller is a pair of integers $(x, y)$
that represents the acceleration vector (on a 2D plane), with $-10 \le x, y \le 10$.
Therefore, the alphabet $\Sigma$ has 441 symbols.
The monitor computes the running average of weights of the automaton $\A_M$
that has one state for each symbol from $\Sigma$, and from each state
$q$ there is an outgoing edge $q \xrightarrow{q' : x} q'$ to any other state $q'$ under the symbol $q'$.
In other words, the states remember the last issued action.
The weight $x$ of each transition going from $q$ to $q'$ is the distance between $q$ and $q'$.
In total, the automaton $\A_M$ has 441 states and 194481 edges.
% Therefore, more extreme changes in the position are assigned bigger weights
% and the running average of the weights is the smaller the smoother (on average) is the
% sequence of actions issued by the controller.

The initial mission of the drone was to get from the point $(0, 0)$ to $(1000, 1000)$
(with no obstacles) using a controller that every 0.1\,s issues a command to
accelerate toward the target. However, a random deviation
taken from the normal distribution with mean 0 and standard deviation either 1 or 5 (this is a parameter) is applied to both acceleration coordinates at every step.
The magnitude of the acceleration is also random, skewed toward the maximum acceleration
value 10. If the resulting acceleration along a coordinate is greater (lower) than 10 (-10, resp.), it is set to 10 (-10, resp.).

The rather chaotic controller described above models an imperfect controller
and results in navigating the drone along an erratic trajectory.
We ran another mission where the drone followed the previously taken erratic trace that has been smoothed using gradient ascent.
The situation is depicted on the left in \cref{tab:drone},
and the results of monitoring the trajectories is on the right in the same figure.
The monitor correctly assigns lower scores to smoother trajectories,
which directly corresponds to the difference in energy consumption (EC).

	\section{Conclusion} \label{sec:conclusion}

We presented QuAK: the first software tool to automate the analysis of quantitative automata.
QuAK implements several standard decision procedures as well as an antichain-based inclusion checking, algorithms to decide whether an automaton is safe, live, and constant, and a construction for its safety-liveness decomposition.
In the future, we plan to add algorithms for discounted sum automata and implement safety-liveness decompositions for more classes of automata.
One can also extend the tool with a support for probabilistic and nested variants of quantitative automata.

%	\newpage
	\bibliographystyle{splncs04}
	\bibliography{main}

\end{document}